\documentstyle[12pt]{article}
\begin{document}
\begin{titlepage}
\begin{center}

$\,$  \vspace{2cm}

{\Large Non-extensive statistical mechanics approach to fully developed
hydrodynamic turbulence}

\vspace{2.cm}
{\bf Christian Beck}

\vspace{1cm}

School of Mathematical Sciences, Queen Mary and Westfield College,
University of London, Mile End Road, London E1 4NS, England
\end{center}

\vspace{3cm}

\abstract{
We apply non-extensive methods to the statistical analysis of fully
developed turbulent flows. Probability density functions of velocity differences
at distance $r$ obtained by extremizing the Tsallis entropies
coincide well with what is
measured in turbulence experiments.
We derive a set of
relations between the hyperflatness
factors $F_m$ and the non-extensitivity parameter $q$, which can
be used to directly extract the function $q(r)$ from experimentally measured
structure functions. We comment on various non-extensive methods
to calculate
the moment scaling exponents $\zeta_m$.}

\vspace{1.3cm}

\end{titlepage}

\section{Introduction}

Consider a suitable observable in a fully
developed hydrodynamic turbulent flow. For example, this may
be a longitudinal
or transverse velocity difference, a temperature or a pressure difference.
The dynamics is effectively described by some highly nonlinear
set of model equations---for example the Navier-Stokes
equation.
Since nobody is able to solve this equation exactly, it is desirable
to find some effective statistical description using
methods from generalized statistical mechanics.

The underlying idea is quite similar to what was going on more than a hundred
years ago, proceeding
from classical mechanics to ordinary thermodynamics. Although nobody was able to
`solve' the classical $N$-body problem with $N\geq 3$ exactly,
one still was able to develop quite a
successful effective theory of a gas of $10^{23}$ particles by extremizing the
Boltzmann-Gibbs entropy.
For fully developed turbulent spatio-temporal chaotic systems, the ordinary
Boltzmann-Gibbs statistics is not sufficient to describe the
(non-Gaussian) stationary state.
But still we can try to develop an effective probabilistic theory using
more general information measures.

The idea to start from an extremization principle
for turbulent flows
is actually not new. For example,
Cocke \cite{poppe} has presented
some work on turbulence where he extremizes the Fisher information.
Castaing et al. \cite{cast}
also start from another extremum principle to derive probability densities in
fully
developed turbulence. 
Here we will work within a new approach to turbulence \cite{hydro, ramos, ari} based on
extremizing the Tsallis entropies \cite{1,2,3}
\begin{equation}
S_q= \frac{1}{q-1} \left( 1- \sum_i p_i^q \right).
\end{equation}
The $p_i$ are the probabilities of the various microstates of the
physical system, and $q$ is the non-extensitivity parameter.
The
ordinary Boltzmann-Gibbs entropy is obtained in the limit
$q \to 1$.

Generally, the Tsallis entropies
are known to have several nice properties. They are positive,
concave, take on their extremum for the uniform distribution and
preserve the Legendre transform structure of thermodynamics.
On the other hand, they
are non-extensive (non-additive for independent subsystems).

Extremizing $S_q$ under suitable
norm and energy constraints, one arrives
at a generalized version of the canonical distribution given by
\begin{equation}
p_i= \frac{1}{Z_q} (1+(q-1)\beta \epsilon_i )^{\frac{1}{1-q}},\label{tsa}
\end{equation}
where
\begin{equation}
Z_q= \sum_i (1-(1-q) \beta \epsilon_i)^{\frac{1}{1-q}}
\end{equation}
is the partition function, $\beta=1/(kT)$ is a suitable
inverse temperature variable, and the $\epsilon_i$ are the
energies of the microstates $i$.  Ordinary thermodynamics is
recovered for $q\to 1$.
One can also work
with the escort distributions \cite{BS}, defined by $P_i=p_i^q/\sum p_i^q$.
If $\beta$ is allowed to depend on $q$, the escort distribution
is of  the same form as eq.~(\ref{tsa}),
with a new $q'$ defined by
$q/(q-1)=:1/(q'-1)$.

All we have to decide now is what we should take for
the effective energy levels $\epsilon_i$ in the
turbulence application. This depends on the problem
considered. For example, turbulence in
different dimensions ought to yield different effective
energy levels. Moreover, different observables will also lead to different
effective energies. In a turbulent 3-dimensional flow for example, temperature differences
should be described by different effective energies than
velocity differences. This is
clear from the fact that the experimentally observed
stationary probability distributions
(slightly) differ.

At this stage one has to turn to some sort of model.
In the following we will
consider a simple local model for longitudinal velocity differences
that seems to reproduce the statistics of true turbulence
experiments quite well  \cite{hydro}.

\section{Perturbative approach to chaotically driven systems}

The model is based on a generalization of the Langevin equation to
deterministic chaotic driving forces \cite{BR}--\cite{18}. We denote the local longitudinal velocity
difference of two points in the liquid
separated by a distance
$r$ by $u$. Clearly $u$
relaxes with a certain damping constant $\gamma$ and at the same
time is driven by deterministic chaotic force differences $F_{chaot}(t)$ in
the liquid, which are very complicated. Hence a very simple local
model is
\begin{equation}
\dot{u}=-\gamma u +F_{chaot} (t) \label{20} .
\end{equation}
The force $F_{chaot}(t)$ is not Gaussian white noise
but a complicated deterministic chaotic forcing. It
changes on a typical time scale $\tau$, which is
smaller than the relaxation time $\gamma^{-1}$,
so $\gamma \tau$ is a small parameter.
We may effectively discretize in time and consider the rescaled kick force
\begin{equation}
F_{chaot}(t)=
(\gamma \tau)^{1/2} \sum_{n=0}^\infty x_n \delta (t-n\tau),
\end{equation}
where the $x_n$ are the iterates of an appropriate
stroboscopic chaotic map $T$.
Integrating eq.~(\ref{20}) one obtains
\begin{eqnarray}
x_{n+1} &=&T(x_n)  \nonumber \\
u_{n+1} &=& \lambda u_n +\sqrt{\gamma \tau} x_{n+1},
\label{skew}
\end{eqnarray}
where $u_n:=u(n\tau +0)$ and
$\lambda :=e^{-\gamma \tau}$. For $\gamma\tau \to 0$,
$t=n\tau$ finite,
and so-called $\varphi$-mixing deterministic maps $T$ it
has been shown \cite{BR} that the $u$-dynamics converges to the
Ornstein-Uhlenbeck process, regarding the initial values $x_0$
as random variables.
 Hence the invariant density of $u$
becomes Gaussian in this limit. For finite $\gamma\tau$, on the
other hand, the invariant density is non-Gaussian. It can have
fractal and singular properties if $\gamma\tau$ is large \cite{18}.
But for small $\gamma\tau$ it approaches the Gaussian distribution
provided $T$ is $\varphi$-mixing.

The route to the Gaussian limit behaviour has been
investigated in detail in \cite{20} for many different chaotic
maps $T$. Within a well defined universality class (defined by square
root scaling of the first order correction) it was found that for small
enough $\gamma\tau$
the invariant density of
$u$ is always given by
\begin{equation}
p(u)=\frac{1}{\sqrt{2\pi}}
e^{-\frac{1}{2}u^2+c\sqrt{\gamma \tau} (u-\frac{1}{3}u^3)}
+O(\gamma \tau ), \label{leverhulme}
\end{equation}
provided the variable $u$ is rescaled such that the
variance of the distribution
is 1. $c$ is a non-universal constant.
This means, not only the Gaussian limit distribution is universal
(i.e.\ independent of details of $T$),
but also the way the Gaussian is approached if the time scale ratio
$\gamma\tau$ goes to 0. Much more details on this can be found in
\cite{20}.

It is now reasonable to assume that also the local chaotic forces
acting on longitudinal
velocity differences in a turbulent flow lie in this
universality class. We  can then use eq.~(\ref{leverhulme})
to construct effective energy levels $\epsilon_i$ for the
non-extensive theory.

\section{Constructing effective energy levels}

Eq.~(\ref{leverhulme}) corresponds to a Boltzmann factor
\begin{equation}
p_i=\frac{1}{Z} e^{-\beta \epsilon_i}
\end{equation}
with $\beta =1/(kT)=1$, $Z=\sqrt{2\pi}$, and
energy $\epsilon_i$ formally given by
\begin{equation}
\epsilon_i=\frac{1}{2}u^2 -c \sqrt{\gamma \tau}
(u-\frac{1}{3}u^3) +O(\gamma\tau ) .\label{en}
\end{equation}
The main contribution is the kinetic energy $\frac{1}{2}u^2$,
but in addition there is also a small asymmetric term with
a universal $u$-dependence.

The complicated hydrodynamic interactions and
the cascade of energy dissipating from larger to smaller levels is now expected
to be effectively described by a non-extensive theory with
the above energy levels (see \cite{PRE} for a related cascade model).
We obtain from eq.~(\ref{tsa}) and (\ref{en}) the formula
\begin{equation}
p(u)=\frac{1}{Z_q}\left( 1+\beta (q-1) \left( \frac{1}{2} u^2-
c\sqrt{\gamma\tau}(u-\frac{1}{3}u^3)+O(\gamma\tau)
 \right) \right)^{-\frac{1}{q-1}} \label{almut}.
\end{equation}
This equation is in very good agreement with experimentally
measured probability densities.
For detailed comparisons with various turbulence experiments, see
\cite{hydro,BLS,bod,sreeni}.
The parameter $\beta$ is determined by the condition that the distribution
should have variance 1. For $\gamma\tau =0$ this is achieved for
$\beta=2/(5-3q)$.

Note that we have identified a small parameter $\gamma\tau$ in our
approach. It is the ratio of two time scales---that of the local
forcing and that of the relaxation to the stationary state. One
may conjecture that it is related to the inverse Reynolds number
$R_\lambda^{-1}$ \cite{hydro}. The turbulent statistics is determined
by a kind of effective non-extensive field theory with the formal
coupling constant $\sqrt{\gamma\tau}$. A perturbative approach is
possible since $\sqrt{\gamma\tau}$ is small. In fact, for the
dynamics (\ref{skew}) one can work with analogues of Feynman
graphs related to higher-order correlations of the chaotic
dynamics \cite{18a,HB}. Eq.~(\ref{almut}) is just obtained by 
first-order perturbation theory---the complete theory is the infinite-order
theory taking into account all orders of $\sqrt{\gamma\tau}$. But
first we have to understand the `free' turbulent field theory
obtained for $\gamma\tau =0$. Here almost everything can be
calculated analytically.

\section{The `free' turbulent field theory}

If  $\gamma\tau =0$
the moments can be easily evaluated. In \cite{hydro} we obtained
\begin{equation}
\langle |u|^m \rangle = \int_{-\infty}^\infty p(u)|u|^m du =
\left( \frac{2k}{\beta} \right)^{\frac{m}{2}}
\frac{ B\left( \frac{m+1}{2}, k- \frac{m+1}{2} \right)}{B \left( \frac{1}{2},
k-\frac{1}{2} \right)}
\end{equation}
with $k$ defined as $k:=1/(q-1)$ ($k$ needs not
to be integer). The beta function is defined as
\begin{equation}
B(x,y)=\frac{\Gamma(x)\Gamma(y)}{\Gamma (x+y)}.
\end{equation}
We now show that this formula for the moments can be significantly simplified.
First, we obtain from the definition of the beta function
\begin{equation}
\langle |u|^m \rangle = \left( \frac{2k}{\beta} \right)^{\frac{m}{2}}
\frac{\Gamma \left( \frac{1}{2}+\frac{m}{2}\right)\Gamma
\left( k-\frac{1}{2}-\frac{m}{2}\right)}{\Gamma \left( \frac{1}{2} \right)
\Gamma \left( k-\frac{1}{2} \right)} \label{be}
\end{equation}
Generally, one has
\begin{equation}
\Gamma (x+n) =\Gamma (x) \prod_{j=0}^{n-1} (x+j)\label{pro}
\end{equation}
for natural numbers $n$. Suppose $\frac{m}{2}=:n \in {\bf N}$,
then it follows from eq.~(\ref{be}) and  (\ref{pro})
\begin{equation}
\langle u^m \rangle =\left(\frac{2k}{\beta} \right)^{\frac{m}{2}}\;
\prod_{j=0}^{\frac{m}{2}-1}\frac{2j+1}{2j+2k-m-1}
\end{equation}
In particular, we obtain for the first few moments
\begin{eqnarray}
\langle u^2 \rangle &=&\left( \frac{2k}{\beta}
\right)\frac{1}{2k-3}
\\ \langle u^4 \rangle &=&\left( \frac{2k}{\beta}\right)^2
\frac{1}{2k-5}\frac{3}{2k-3}\\ \langle u^6 \rangle &=&\left(
\frac{2k}{\beta}\right)^3
\frac{1}{2k-7}\frac{3}{2k-5}\frac{5}{2k-3}\\
\end{eqnarray}
This can be used to evaluate the complete set of hyperflatness factors $F_m$
defined as
\begin{equation}
F_m=\frac{\langle u^{2m}\rangle }{\langle u^2\rangle ^m}
\end{equation}
We obtain
\begin{eqnarray}
F_1 &=& 1 \\
F_2 &=& 3 \; \frac{2k-3}{2k-5}\label{21} \\
F_3 &=& 3\cdot 5 \; \frac{(2k-3)^2}{(2k-7)(2k-5)}  \label{22}   \\
F_4 &=& 3\cdot\ 5\cdot 7 \; \frac{(2k-3)^3}{(2k-9)(2k-7)(2k-5)}
\end{eqnarray}
and generally
\begin{equation}
F_m=(2m-1)!! \frac{(2k-3)^{m-1}}{\prod_{j=5}^{2m+1}(2k-j)}
\end{equation}
($j$ odd). Note that all hyperflatness factors are independent of $\beta$.

\section{Extracting $q(r)$ from experimentally measured structure functions}

The great
advantage of the hyperflatness factors is that they yield  a direct way
to estimate the $r$-dependent non-extensitivity parameter $q(r)$ from
experimentally measured structure
functions $\langle u^m \rangle (r)$.  Eq.~(\ref{21})
yields
\begin{equation}
F_2=\frac{6k-9}{2k-5}=\frac{15-9q}{7-5q}
\end{equation}
or
\begin{equation}
k=\frac{5F_2-9}{2F_2-6},
\end{equation}
equivalent to
\begin{equation}
q=\frac{7F_2-15}{5F_2-9}
\end{equation}
(see also \cite{ramos}).

Another relation follows from  eq.~(\ref{22}).
 \begin{equation}
k=\frac{1}{2F_3-30} \left( 6F_3-45\pm\sqrt{F_3^2+120F_3}\right),
\end{equation}
equivalent to
\begin{equation}
q=\frac{8F_3-75\pm\sqrt{F_3^2+120F_3}}{6F_3-45\pm\sqrt{F_3^2+120F_3}}
\end{equation}
In fact, each hyperflatness factor $F_m$ with $m\geq 2$
yields a relation for $k$ (or $q$), and
all relations are the same in case the `free' turbulence theory is exact.

Given some experimentally measured
hyperflatness structure functions $F_m(r)$ one can now determine the
corresponding curves $q(r)$. The less these curves differ for the various $m$, the
more precise is the zeroth-order (free) turbulence theory.
For examples of such experimentally measured curves $q(r)$,
see \cite{BLS}. Generally, the hyperflatness factors
are complicated functions of both the Reynolds number $R_\lambda$
and the separation distance $r$, and
so is $q(R_\lambda , r)$. But one may conjecture that for
$R_\lambda \to \infty$ one obtains a universal function $q^*(r)$.
How it looks like in the entire $r$-range is still an open question.

\section{Some remarks on the scaling exponents $\zeta_m$}

The precise values of the scaling exponents $\zeta_m$, which
describe the scaling behaviour of the
structure functions $\langle u^m \rangle \sim r^{\zeta_m}$
in the inertial range,
are still a rather controversial topic
 in  turbulence theory. In fact, it is not at all
clear whether there is exact scaling at all or just
approximate scaling, whether the higher moments exist at all
or not,  how reliable the experimentally measured
higher-order exponents are, and what the effects of finite
Reynolds numbers are. On the theoretical side,
a variety of models and theories have been suggested
(see, e.g., \cite{frisch, bohr})
but a true breakthrough convincing a significant majority
of  scientists working in the field seems not to have been
reached at the moment.

To `derive' values for the scaling exponents using methods
from non-extensive
statistical mechanics, one needs additional assumptions---
just as in all other models and theories dealing with
the scaling exponents. In \cite{hydro} a logarithmic
depence of  $k$ on $r$ was suggested  to derive
the following formula
for the scaling
exponents
\begin{equation}
\zeta_m=\frac{m}{3} \left\{ 1-\log_2 \left( 1-\frac{3}{2\bar{k}-1}\right)
\right\}+\log_2\left( 1-\frac{m}{2\bar{k}-1}\right). \label{zeta1}
\end{equation}
Here $\bar{k}=1/(q-1)$ is the average value of the non-extensitivity
parameter in the inertial range. On the other hand,
Kolmogorov's lognormal model
(the K62-theory \cite{K62}) predicts
\begin{equation}
\zeta_m=\frac{m}{3} \left( 1+\frac{\mu}{2}\right)-\frac{1}{18}\mu m^2.
\label{zeta2}
\end{equation}
where $\mu$ is the intermittency parameter. Now, asuming
that $\bar{k}$ is rather large we can expand the logarithms
in eq.~(\ref{zeta1}).
The linear terms cancel, and
the first non-trivial terms are the quadratic ones. Neglecting
the higher-order cubic terms, one precisely obtains from
eq.~(\ref{zeta1}) the result (\ref{zeta2}), identifying
\begin{equation}
\mu =\frac{18}{\ln 2 (2\bar{k}-1)^2}.
\end{equation}
It is encouraging that the simplest non-extensive
model assumptions lead
to an old theory that has a long tradition in turbulence theory
(though it is known that K62 cannot be correct ---it is,
however, a
good approximation for $m$ not too large).
The value of the intermittency parameter comes out with
the correct order of magnitude ($\mu\approx 0.2$)
if one chooses  $\bar{k}\approx 6$, and the same $k$ fits the
experimentally measured probability densities correctly.

On the other hand, starting in the
derivation leading to eq.~(\ref{zeta1})
from a different dependence of $k$ on $r$ rather than the
logarithmic one,
one also 
obtains different formulas for the scaling exponents. So the final
answer to the question
of the scaling exponents is still open. 

Actually, in \cite{hydro}
eq.~(\ref{zeta1}) was derived using the `free' turbulence
theory, but clearly
the correct theory
is the infinite-order theory. As one can easily see,
the infinite-order theory yields probability densities $p(u)$ living
on a compact support for arbitrary small but finite
$\gamma\tau$.  Indeed, iterating   eq.~(\ref{skew})
one obtains  $u_n=\lambda^nu_0+\sqrt{\gamma\tau}\sum_{j=1}^n
\lambda^{n-j}x_j$. Hence, for any chaotic dynamics $T$ bounded on some
finite phase space (say [-1,1]) one obtains for  $n\to \infty$ 
from the geometric series the
rigorous bound
$|u|\leq \sqrt{\gamma \tau}/(1-\lambda )\sim 1/\sqrt{\gamma\tau}$.
Thus for the infinite-order theory 
{\em all} moments exist for arbitrary small but finite 
$\gamma\tau$ ---in contrast to the 0-th order theory, where
only moments $\langle u^m \rangle$ with $m<2k-1$ exist,
since here the densities live on a non-compact support and there
is polynomial decay for large $|u|$.
This once again shows how delicate the problem of the existence
of the moments and of the scaling exponents in general is.
For densities living on a compact support, as provided by the
infinite-order theory, one expects a linear
asymptotics of  $\zeta_m$ for large $m$.

Generally, the way non-extensive statistics can be used to
`derive' the scaling exponents $\zeta_m$ is not unique. An
alternative approach, based on an extension of the multifractal
model and $q$-values smaller than 1, has been suggested by
T.\ and N.\ Arimitsu \cite{ari}. In their model asymptotically there is a
logarithmic dependence og $\zeta_m$ on $m$.

\section*{Acknowledgement}

A large part of this research was performed during the author's stay
at
the Institute for Theoretical Physics, University of California
at Santa Barbara, supported in part by the National Science
Foundation under Grant No. PHY94-07194. 
The author also gratefully acknowledges support by a
Leverhulme Trust Senior Research Fellowship of the Royal Society.

\end{document}